\documentclass[aps,preprint,superscriptaddress,groupedaddress]{revtex4}  % for double-spaced preprint
\usepackage{graphicx}  % needed for figures
\usepackage{dcolumn}   % needed for some tables
\usepackage{bm}        % for math
\usepackage{amssymb,amsmath}   % for math
\usepackage{verbatim}
\usepackage[T1]{fontenc}
\usepackage{amsmath,amssymb,amsfonts}
 \usepackage{hyperref}
\hypersetup{colorlinks=true,linkcolor=magenta,anchorcolor=green,citecolor=cyan,filecolor=black,menucolor=black,urlcolor=brown}

\usepackage{cleveref}               % if needed

\usepackage{ifpdf}
\usepackage{color}

\usepackage{amsmath}
\usepackage{graphics}
\usepackage{mathtools}
\usepackage[usenames,dvipsnames]{xcolor}
\usepackage{epsfig}
\usepackage{epstopdf}
\usepackage{dcolumn}
\usepackage{tikz}
\usetikzlibrary{shapes.geometric, arrows}
\usepackage{upgreek}
\usepackage{setspace}
\usepackage{enumitem}
\usepackage{array,multirow,bigdelim}

\def\be{\begin{equation}}
\def\ee{\end{equation}}
\def\ba{\begin{eqnarray}}
\def\ea{\end{eqnarray}}

\def\CP1{\mathbb{CP}^1}
\def\SL2C{\mathrm{SL}(2,\mathbb{C})}

\def\Z2{\mathbb{Z}_2}

\def\su2{{SU(2)}}

\def\[{\left[}
\def\]{\right]}

\def\({\left(}
\def\){\right)}
\def\[{\left[}
\def\]{\right]}

\def\i2{\frac{i}{2}}

\def\2F1{\,_2{\rm F}_1}

\newcommand{\beq}{\begin{equation}}
\newcommand{\eeq}{\end{equation}}
\newcommand{\beqq}{\begin{equation*}}
\newcommand{\eeqq}{\end{equation*}}
\newcommand\beqa{\begin{eqnarray}}
\newcommand\eeqa{\end{eqnarray}}
\newcommand\beqaa{\begin{eqnarray*}}
\newcommand\eeqaa{\end{eqnarray*}}
\newcommand\bea{\begin{array}}
\newcommand\eea{\end{array}}

\begin{document}
% The following information is for internal review, please remove them for submission
%\leftline{Version xx as of \today}
%\leftline{Primary authors: Joe E. Physics}
%\leftline{To be submitted to (PRL, PRD-RC, PRD, PLB; choose one.)}
%\leftline{Comment to {\tt d0-run2eb-nnn@fnal.gov} by xxx, yyy}
%\centerline{\em D\O\ INTERNAL DOCUMENT -- NOT FOR PUBLIC DISTRIBUTION}

% the following line is for submission, including submission to the arXiv!!
%\hspace{5.2in} \mbox{Fermilab-Pub-04/xxx-E}

\title{On Post-Minkowskian Hamiltonians in General Relativity}
\author{Andrea Cristofoli}
\affiliation{\smallskip Niels Bohr International Academy and Discovery
  Center\\ Niels Bohr Insitute, University of Copenhagen}
\author{N.E.J. Bjerrum-Bohr}
\affiliation{\smallskip Niels Bohr International Academy and Discovery
  Center\\ Niels Bohr Insitute, University of Copenhagen}
\author{Poul  H. Damgaard}
\affiliation{\smallskip Niels Bohr International Academy and Discovery
  Center\\ Niels Bohr Insitute, University of Copenhagen}
\author{Pierre Vanhove}
\affiliation{Institut de Physique Th\'{e}orique, Universit\'{e}
  Paris-Saclay\\CEA, CNRS, F-91191 Gif-sur-Yvette Cedex, France}
\affiliation{National Research University Higher School of Economics,
  Russian Federation}
\date{\today}
\preprint{SAGEX-19-06-E, IPhT-T/18/172}

\begin{abstract}
We describe the computation of post-Minkowskian Hamiltonians in General Relativity from scattering amplitudes. 
Using a relativistic Lippmann-Schwinger equation, we relate perturbative amplitudes of massive scalars coupled to gravity to the 
post-Minkowskian Hamiltonians of classical General Relativity to any order in Newton's constant. 
We illustrate this by deriving a Hamiltonian for 
binary black holes without spin up to 2nd order in the post-Minkowskian expansion and demonstrate explicitly the 
equivalence with the recently 
proposed method based on an effective field theory matching.
\end{abstract}
%\pacs{04.60.-m, 04.62.+v, 04.80.Cc}
\maketitle

\section{Introduction}
The detection of gravitational waves by the LIGO/Virgo collaboration has opened up the exciting possibility of testing Einstein's theory of 
general relativity at a new and unprecedented level, including the regime of strong gravity as probed by black holes just prior to merging. 
A combination of Numerical Relativity and analytical methods is needed in order to push theory to the level where it can 
provide best-fit templates from which physical parameters can be
extracted. This has spurred interest in 
new and innovative ideas that can facilitate computations of the required two-body interaction Hamiltonians to high accuracy.

Conventionally, the calculations of effective interaction Hamiltonians have been carried out in the systematic post-Newtonian expansion
of General Relativity. The problem can, however, be 
attacked from an entirely different angle, that of relativistic scattering amplitudes as computed by standard quantum field
theory methods in a quantum field theory of gravity coupled to matter \cite{Iwasaki:1971vb}. Modern methods of amplitude
computations greatly facilitate this program \cite{Neill:2013wsa,Bjerrum-Bohr:2013bxa,Vaidya:2014kza,Cachazo:2017jef,Guevara:2017csg,Bjerrum-Bohr:2018xdl,Cheung:2018wkq,Bern:2019nnu}. Incoming and outgoing particles in the scattering process are taken to past and 
future infinity where the metric
by definition is flat Minkowskian, and the full metric is treated perturbatively around that Minkowskian background. The classical
piece of the scattering amplitude solves the scattering problem of two black holes to the given order in Newton's constant $G_N$.
When expanding to the
appropriate post-Newtonian order and defining the interaction potential with the inclusion of the required lower-order Born subtractions as explained in detail in the next section,
the amplitude also contains all the information of the bound state problem of two massive objects to the given order in the expansion 
in Newton's constant. For the bound-state regime one has, on account of the virial theorem, a double expansion in both Newton's
constant and momentum. However, a more daring angle of attack is to treat the bound state problem as not expanded in  momentum while still expanding to fixed order in Newton's constant. Such an approach has recently been proposed by Cheung, Rothstein and
Solon~\cite{Cheung:2018wkq}, and it has already been pushed one order higher in the expansion in Newton's constant~\cite{Bern:2019nnu} 
(and compared to the post-Newtonian expansions in~\cite{Antonelli:2019ytb}). Here the method of effective field theory is used to
extract the interaction Hamiltonian: the underlying Einstein-Hilbert action coupled to matter produces the
classical part of the scattering amplitude while an effective theory of two massive objects define the interaction Hamiltonian.
The correct matching between the two theories is performed by insisting that the two theories produce the same scattering amplitude to
the given order in Newton's constant.

The post-Newtonian expansion (see, $e.g.$, refs.~\cite{Blanchet:2013haa,Levi:2018nxp,Cristofoli:2018bex} for recent comprehensive reviews) of General Relativity
dates back to the founding days of the theory. Its perturbation theory is ideal for the low-velocity situations of planetary orbits,
satellites, and large-distance effects of General Relativity that occur at velocities far below the speed of light. In contrast to
this, the computation of observables in General Relativity based on relativistic scattering amplitudes is valid for all velocities
and in particular this is the proper framework for high energy scattering where kinetic energies can exceed potential energies
by arbitrarily large amounts. This leads naturally to what has become known in the theory of General Relativity
as the post-Minkowskian expansion~\cite{Westpfahl,Ledvinka:2008tk,Damour:2016gwp,Bini:2018ywr,Blanchet:2018yvb,Damour:2017zjx,Guevara:2018wpp}.

Extracting the interaction 
energy from the relativistic scattering amplitude, for consistency with the virial theorem in the bound-state problem 
one would perform a double expansion 
where velocity $v$ and $G_N$ are both kept to the appropriate order. To any given order in $G_N$ this
would imply a truncation of a Taylor-expanded amplitude in powers of momenta. There is no general argument for whether 
keeping higher powers of only one of the expansion parameters in the regime where they are of comparable magnitude will increase
the accuracy. Considering its potential impact, it is nevertheless of much interest to explore the consequences of
keeping higher-order terms of momenta even in the bound state regime where they would not ordinarily
have been included~\cite{Cheung:2018wkq,Bern:2019nnu,Antonelli:2019ytb}. We will here show how that post-Minkowskian
Hamiltonian also follows directly from the full relativistic amplitude without having to perform the amplitude matching to the effective field theory, thereby explicitly showing equivalence between the two approaches~\cite{Bjerrum-Bohr:2018xdl,Cheung:2018wkq}.

\section{Perturbative Gravity as a Field Theory}
\label{sec:headings}
We start by introducing the Einstein-Hilbert action minimally coupled to massive scalar fields $\phi_{a}$
%%%
\begin{equation}
\mathcal{S}=\int d^{4}x \sqrt{-g} \bigg[ \frac{R}{16 \pi G_N}+
\frac{1}{2}\sum_{a}\Big(g^{\mu\nu}\partial_\mu\phi_a\,\partial_\nu\phi_a -m_a^2\phi^2_a\Big)\bigg]\,, \label{action}
\end{equation}
%%%
%%%
where $R$ defines the Ricci scalar and $g\equiv\det(g_{\mu\nu})$.
Perturbatively, we expand the metric around a Minkowski background: $g_{\mu\nu}(x) = \eta_{\mu\nu} + \sqrt{32\pi G_N} h_{\mu\nu}(x)$.
At large distances we can treat the scattering of two massive objects $m_a$ and $m_b$ as that of two point-like particles with the same masses.
This has all been well elucidated in the literature (see, $e.g.$, refs.~\cite{Donoghue:1994dn,BjerrumBohr:2002kt}), although most focus
until now seems to have been on considering the quantum mechanical effects. The way classical terms appear from the
quantum mechanical loop expansion is subtle
\cite{Iwasaki:1971vb,Holstein:2004dn}; see ref.~\cite{Kosower:2018adc} for a very nice and clear discussion of this issue. Instead
of expanding the action (\ref{action}) in terms of ordinary Feynman rules, it pays to use modern amplitude methods to extract
the needed non-analytic pieces in momentum transfer $\vec{q}$ through the appropriate cuts at loop level 
\cite{Neill:2013wsa,Bjerrum-Bohr:2013bxa,Vaidya:2014kza}.

The scattering $m_a+m_b \rightarrow m_a+m_b$ mediated by gravitons at an arbitrary loop order is described by
\begin{equation}
  \mathcal{M}=\begin{gathered}
    \includegraphics[width=3cm]{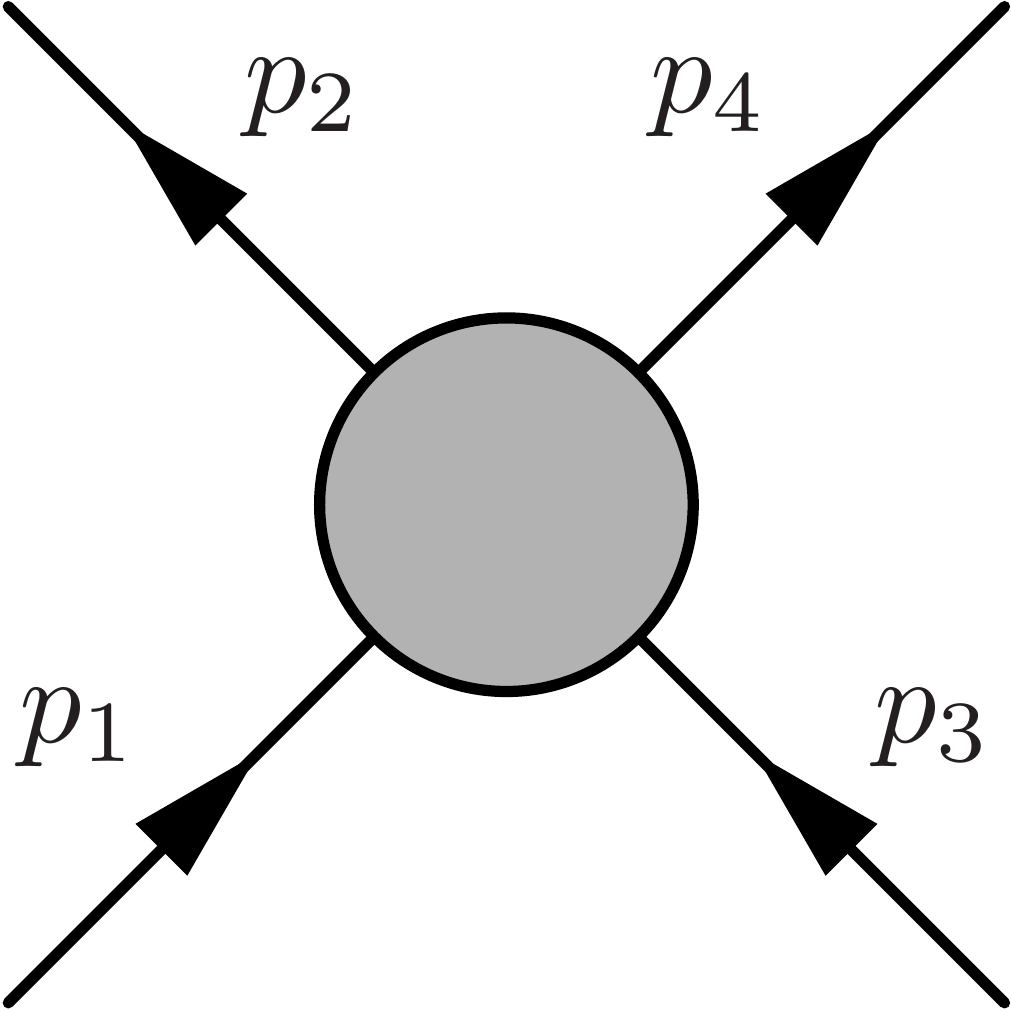}
\end{gathered}
=\sum_{L=0}^{+ \infty} \mathcal{M}^{\rm L- loop}  \quad , \quad \mathcal{M}^{\rm L-loop} \sim O(G_N^{L+1}),
\end{equation}
We choose the center-of-mass frame and parametrize the momenta as follows:
\begin{equation}\begin{split} 
p^{\mu}_{1}&=(E_{a},\vec{p}\,)\,, \  \ \, p^{\mu}_{2}=(E_{a},\vec{p}\,')\,, \\ 
p^{\mu}_{3}&=(E_{b},-\vec{p}\,)\,, \  p^{\mu}_{4}=(E_{b},-\vec{p}\,')\,,
\end{split}\end{equation}
and $|\vec{p}\,|=|\vec{p}\,'|$. We also define
\begin{equation}
q^{\mu}=p^{\mu}_{1}-p^{\mu}_{2}=p^{\mu}_{4}-p^{\mu}_{3}\equiv(0,-\vec{q}\,)\,, \ \ \vec{q}\equiv\vec{p}\,'-\vec{p}\,,
\end{equation}%%%
and the total energy $E_p=E_a+E_b$.
%%%

\section{The Lippmann-Schwinger Equation} 

It is a classical problem in perturbative scattering theory to relate the scattering amplitude ${\mathcal M}$ to an interaction 
potential $V$. This is typically phrased in terms of non-relativistic quantum mechanics, but it is readily generalized
to the relativistic case. Crucial in this respect is the fact that we shall consider particle solutions to the relativistic equations only.
There will thus be, in the language of old-fashioned (time-ordered) perturbation theory, no back-tracking diagrams corresponding
to multiparticle intermediate states. This is trivially so since we neither wish to treat the macroscopic classical objects
such as heavy neutron stars as indistinguishable particles with their corresponding antiparticles, nor do we wish to probe
the scattering process in any potential annihilation channel. The classical objects that scatter will always be restricted
to classical distance scales.

We now briefly outline a systematic procedure for connecting the scattering amplitude in perturbative gravity 
with post-Minkowskian potentials in classical General Relativity. We start by introducing a bit of notation.
First, we assume the existence of a relativistic one-particle Hamiltonian of only particle states describing what
in bound-state problems is known as the Salpeter equation,
\begin{equation}
\hat{\mathcal{H}}=\hat{\mathcal{H}}_{0}+\hat{V}, \label{formalH} \qquad \mathcal{\hat{H}}_{0}=\sqrt{\hat{k}^2+m^2_a}+\sqrt{\hat{k}^2+m_b^2}
\end{equation}
where $\hat{V}$ is a so far unspecified potential describing our post-Minkowskian system. 
We also define, on a proper subset of the complex plane, the following $\mathbb{C}$-valued operators
\begin{equation}
\hat{G}_0(z) \equiv (z-\hat{\mathcal{H}_0})^{-1}, \qquad \hat{G}(z) \equiv (z-\hat{\mathcal{H}})^{-1},
\end{equation}
\begin{equation}
\hat{T}(z) \equiv \hat{V}+\hat{V}\hat{G}(z)\hat{V} \label{Gdef}
\end{equation}
Here $\hat{G}_0$ and $\hat{G}$ are the Green's operator for the free and interacting case, while $\hat{T}$ is the off-shell
scattering matrix whose on-shell matrix elements provide the non-trivial $S$-matrix elements. We can relate the two Green's operator by means of the following operator identity
\begin{equation}
A^{-1}=B^{-1}+B^{-1}(B-A)A^{-1} \quad \Rightarrow \quad \hat{G}=\hat{G}_0+\hat{G}_0\hat{V}\hat{G} \label{gi}
\end{equation}
Multiplying both sides of (\ref{Gdef}) by $\hat{G}_0$, combined with ($\ref{gi}$), one has
\begin{equation}
\hat{G}_0\hat{T}= \hat{G}_0\hat{V}+\hat{G}_0\hat{V}\hat{G}\hat{V}=\hat{G}\hat{V},
\end{equation}
\begin{equation}
\hat{T}(z)=\hat{V}+\hat{V}\hat{G}_0(z)\hat{T}(z)
\end{equation}
which is the basis for a perturbative knowledge of $\hat{T}$ and it usually known as \emph{Lippmann-Schwinger equation}.\\
We now take the inner product on scattering states $|p\rangle, |p' \rangle$  
\begin{equation}
\langle p|\hat{T}(z)|p'\rangle=\langle p | \hat{V}|p'\rangle+\int \frac{d^3k}{(2 \pi)^3} 
\frac{\langle p| \hat{V}|k\rangle \langle k|\hat{T}(z)|p' \rangle}{z-E_k} \label{Integraleq}
\end{equation}
and use the crucial relation
\begin{equation}
\label{niu}
\lim_{\epsilon \rightarrow 0} \langle p|\hat{T}(E_p+i \epsilon)|p'\rangle = \mathcal{M}(p,p')
\end{equation} 
which provides the link to the conventionally defined scattering amplitude $\mathcal{M}$ in quantum field theory restricted to the particle sector. Substituting (\ref{niu}) into (\ref{Integraleq}) we have a recursive relation between the amplitude and the post-Minkowskian potential
\begin{equation}
\mathcal{M}(p,p')=\langle p|V|p'\rangle+\int \frac{d^3k}{(2 \pi)^3}
\frac{ \langle p|V|k\rangle \mathcal{M}(k,p')}{E_p-E_k+i \epsilon}\,
\end{equation}
Solving this equation iteratively, we can invert it in order to arrive at a relativistic equation for the potential $V$
\begin{equation}
\langle p|V|p'\rangle=\mathcal{M}(p,p')-\int \frac{d^3k}{(2 \pi)^3}
\frac{\mathcal{M}(p,k) \mathcal{M}(k,p') }{E_p-E_k+i \epsilon}+\cdots
\label{VBorn}
\end{equation}
or, in position space,
\begin{equation}
\label{vi}
V(p,r)=\int \frac{d^3q}{(2 \pi)^3}e^{i q \cdot r} V(p,q), 
\end{equation}
with
\begin{equation}
V(p,q)  \equiv \langle p|V|p'\rangle
\end{equation}
At this stage there has not been any restriction to a non-relativistic
limit. The anti-particle sector has been eliminated by hand, as dictated by the physical scattering process.
We can thus regard (\ref{vi}) as defining a post-Minkowskian potential.

\section{Post-Minkowskian Hamiltonians}
\subsection{The post-Minkowskian potential to first order}
We are now ready to use the above definition of the relativistic interaction potential to describe the post-Minkowskian Hamiltonian to the trivial lowest order for two massive scalars of masses $m_a$ and $m_b$ interacting with gravity. With the non-relativistic normalization of external states,
\begin{equation}
\mathcal{M}^{\rm tree}=\begin{gathered}
\includegraphics[width=3cm]{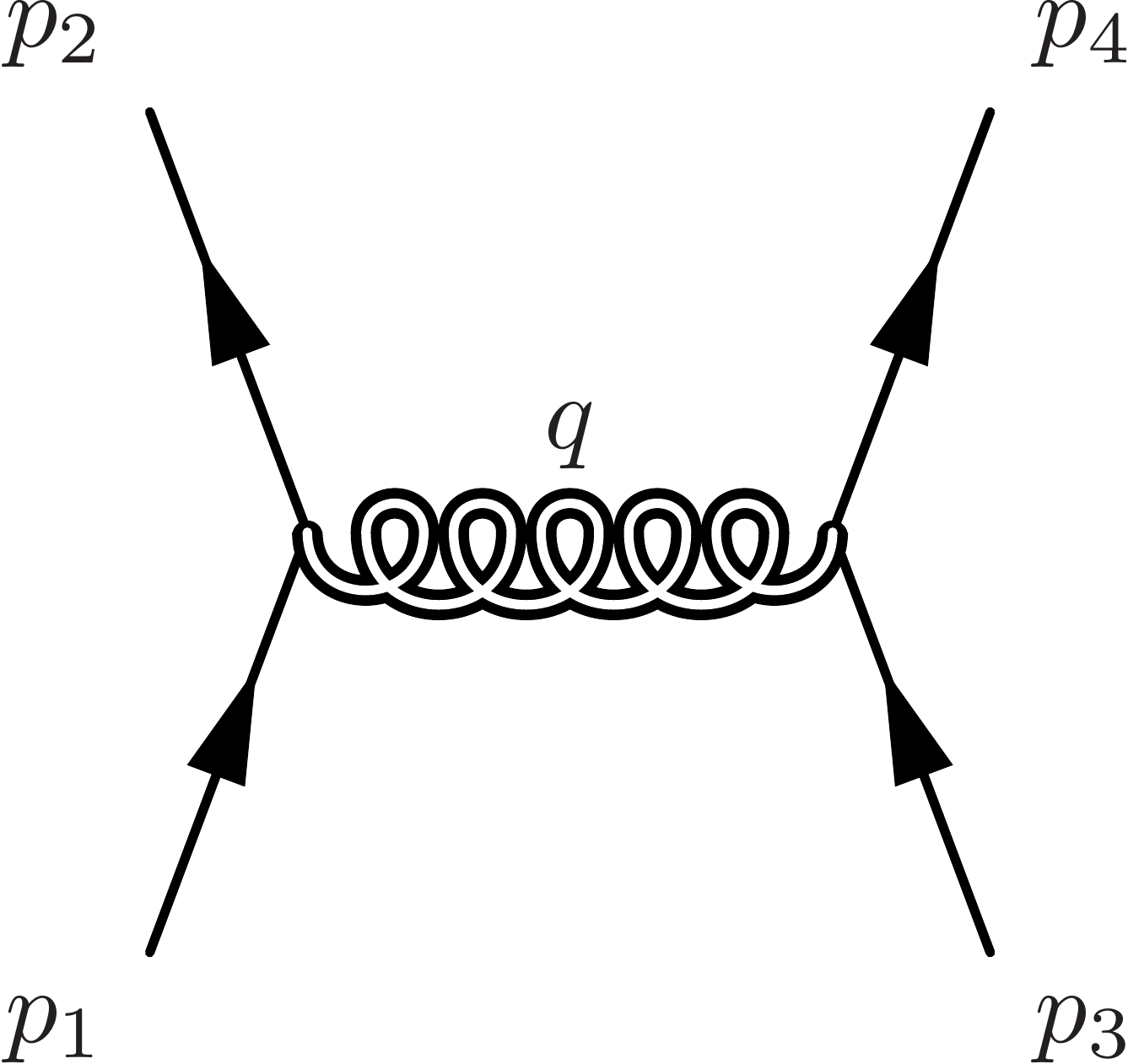}
\end{gathered}
=
\frac{4 \pi G_N}{\sqrt{E_a(p_1) E_a(p_2)E_b(p_3) E_b(p_4)}}\frac{A(p_1,p_2,p_3,p_4)}{q^2},
\end{equation}
with
\begin{equation}
A(p_1,p_2,p_3,p_4)=(p_1\cdot p_3) (p_2 \cdot p_4) + (p_1 \cdot p_4) (p_2 \cdot p_3)- (p_1 \cdot p_2) (p_3 \cdot p_4) + (p_1 \cdot p_2) m_b^2 + (p_3 \cdot p_4) m_a^2 -2 m_a^2 m_b^2
\label{Atree}\end{equation}
In the center-of-mass frame this reduces to
\begin{equation}
\mathcal{M}^{\rm tree}=-\frac{4\pi G_N}{E_a E_b}\frac{[2(p_1\cdot p_3)^2-m^2_am^2_b-|\vec{q}\,|^2(p_1 \cdot p_3)]}{|\vec{q}\,|^2},
\end{equation} 
with $p_1 \cdot p_3= E_a(p) E_b(p)+|\vec{p}\,|^2$.\\
In order to facilitate a comparison with~\cite{Cheung:2018wkq} we can write the Fourier transform as 
\begin{equation}
V_{ 1PM}(p,r)=\frac{1}{E^2_p \xi}\frac{G_Nc_1(p^2)}{r} + \cdots \label{Vtree}
\end{equation}
with
\begin{equation}
c_1(p^2) \equiv m^2_am^2_b-2(p_1 \cdot p_3)^2 \quad , \quad \xi \equiv \frac{E_a E_b}{E_p^2} ~ \label{c1def}
\end{equation}
The terms omitted in eq. (\ref{Vtree}) are either ultra-local or vanishing in the classical limit. This of course agrees with
the leading-order potential of ref.~\cite{Cheung:2018wkq} while not very easily derived in more traditional approaches.

\subsection{The post-Minkowskian potential to second order}
In order to consider a post-Minkowksian potential at second order in $G_N^2$, we will need to consider a contribution coming from the iterated tree-level amplitude, as dictated by (\ref{VBorn})
\vspace{2mm}
\begin{equation}
\label{w}
V_{\rm 2PM}(p,q)=\mathcal{M}^{\rm 1-loop}(p,p')+\mathcal{M}^{\rm Iterated}(p,p'),
\end{equation}
\begin{equation}
\mathcal{M}^{\rm Iterated}(p,p') \equiv -\int \frac{d^dk}{(2 \pi)^d}
\frac{\mathcal{M}^{\rm tree}(p,k) \mathcal{M}^{\rm tree}(k,p') }{E_p-E_k+i \epsilon} ~.
\end{equation}
\newline
Infrared divergences are regularized by temporarily switching to $d+1$ space-time dimensions.
The classical terms of the one-loop amplitude have been given elsewhere~\cite{Bjerrum-Bohr:2018xdl,Cheung:2018wkq,Holstein:2008sx,Guevara:2017csg,KoemansCollado:2019ggb,Brandhuber:2019qpg,Emond:2019crr}. They can be decomposed in terms of scalar integrals with coefficients that are independent of the exchanged three-momentum $\vec{q}$,
\begin{equation}
\label{total}
\mathcal{M}^{\rm 1-loop}=\frac{i16 \pi^2
    G_N^2}{E_aE_b}\bigg(c_{\Box}\mathcal{I}_{\Box}+ c_{\bowtie} \mathcal{I}_{ \bowtie}+
c_{\triangleright}\mathcal{I}_{\triangleright}+c_{\triangleleft}\mathcal{I}_{\triangleleft}+\cdots \bigg)
\end{equation}
where the symbol of each coefficient refers to the topology of the contributions involved while the ellipses denote quantum mechanical contributions that we neglect.\\
In detail, the scalar box and crossed-box integrals are given by
\vspace{2mm}
\begin{align}
  \mathcal{I}_{\Box} &=\!\!\int\! \! \frac{d^{d+1}\ell}{(2
                       \pi)^{d+1}}\frac{1}{((\ell+p_1)^2-m_a^2+i\varepsilon)((\ell-p_3)^2-m_b^2+i\varepsilon)(\ell^2+i\varepsilon)((\ell+q)^2+i\varepsilon)},\\
  \mathcal{I}_{\bowtie} &=\!\!\int\! \! \frac{d^{d+1}\ell}{(2
                       \pi)^{d+1}}\frac{1}{((\ell+p_1)^2-m_a^2+i\varepsilon)((\ell+p_4)^2-m_b^2+i\varepsilon)(\ell^2+i\varepsilon)((\ell+q)^2+i\varepsilon)}\,,
\end{align}
\newline
At leading order in the momentum transfer $\vec q$ the coefficients of these integrals are finite at 
$d=3$ and given by~\cite{BjerrumBohr:2002kt,Bjerrum-Bohr:2013bxa} 
\begin{equation}
  c_{\Box}=c_{\bowtie}= 4 \big(m^2_am^2_b-2(p_1 \cdot p_3)^2\big)^2 ~.
\end{equation}
The scalar triangle integrals are given by
\begin{align}
\mathcal{I}_{\triangleright}&=\int \frac{d^{d+1}\ell}{(2
  \pi)^{d+1}}\frac{1}{((\ell+q)^2+i\varepsilon)(\ell^2+i\varepsilon)((\ell+p_1)^2-m^2_a+i\varepsilon)} \\
  \mathcal{I}_{ \triangleleft}&=\int \frac{d^{d+1}\ell}{(2
  \pi)^{d+1}}\frac{1}{((\ell-q)^2+i\varepsilon) (\ell^2+i\varepsilon) ((\ell-p_3)^2-m_b^2+i\varepsilon)}
\end{align}
with coefficients, at the leading order in $|\vec q|$ and around $d=3$, given by
\begin{equation}
c_{\triangleright}=3 m^2_{a}\big(m^2_am^2_b-5(p_1
\cdot p_3)^2\big) \quad , \quad   c_{ \triangleleft}=3 m^2_{b}\big(m^2_am^2_b-5(p_1
\cdot p_3)^2\big)
\end{equation}
These scalar integrals are conveniently evaluated by performing proper contour integrals  in $\ell^0$ as explained in~\cite{Bjerrum-Bohr:2018xdl}. Doing so, we see that the box, crossed-box, and triangle contributions are given by~\cite{BjerrumBohr:2002kt,Donoghue:1996mt}
\begin{equation}
  \mathcal{I}_{\Box}
                    =-\frac{i}{16\pi^2|\vec{q}\,|^2}\left(-\frac{1}{m_am_b}+{m_a(m_a-m_b)\over
                      3m_a^2m_b^2}+{i\pi \over |p|E_p}\right)\ \bigg(\frac{2}{3-d} -\log|\vec{q}\,|^2\bigg)+\cdots,
                  \end{equation}
\begin{equation}
  \mathcal{I}_{\bowtie}
                     =-\frac{i}{16\pi^2|\vec{q}\,|^2}\left(\frac{1}{m_am_b}-{m_a(m_a-m_b)\over
                      3m_a^2m_b^2}\right)\bigg(\frac{2}{3-d} -\log|\vec{q}\,|^2\bigg)+\cdots,
\end{equation}                  
\begin{equation}
  \mathcal{I}_{\triangleright}=-\frac{i}{32
 m_a }\frac{1}{|\vec{q}\,|}+\cdots,
 \end{equation}
 \begin{equation}
 \mathcal{I}_{\triangleleft}=-\frac{i}{32
  m_b}\frac{1}{|\vec{q}\,|}+\cdots \,,
\end{equation}
at leading order in the ${|\vec q\,|}^2$ expansion and around $d=3$.
We thus arrive at the one-loop amplitude to leading order in $|\vec{q}\,|^2$,
\begin{equation}
\label{uo}
\mathcal{M}^{\rm 1-loop}=\frac{\pi^2 G_N^2}{E^2_p \xi}\bigg[\frac{1}{2|\vec{q}\,|}
\bigg(\frac{c_{\triangleright}}{m_{a}}+\frac{c_{\triangleleft}}{m_b}\bigg)+
\frac{i}{E_p}\frac{c_{\Box}}{|\vec{p}\,|}\frac{(\frac{2}{3-d}-\log|\vec{q}\,|^2)}{\pi|\vec{q}\,|^2}\bigg]\,
\end{equation}
The imaginary part of this which arises from the box and crossed-box integrals is the infrared divergent Weinberg phase
\cite{Weinberg:1965nx}. By restoring the $\hbar$-counting, one sees that it scales as $\hbar^{-1}$, 
a behavior dubbed super-classical in  ~\cite{Kosower:2018adc}.
We will show below that it cancels in the properly defined potential, a fact already noted in the 
post-Newtonian expansion~\cite{Holstein:2008sx}.

We next evaluate the iterated tree-level contribution given by
\begin{equation}
\label{Buorn}
\mathcal{M}^{\rm Iterated}=-\frac{16 \pi^2 G_N^2}{E_a(p^2)E_b(p^2)} \int \frac{d^d k}{(2\pi)^d}
\frac{A(\vec{p},\vec{k})}{|\vec{p}-\vec{k}|^2}\frac{A(\vec{k},\vec{p}\,')}{|\vec{p'}-\vec{k}|^2}\frac{\mathcal{G}(p^2,k^2)}{E_a(k^2) E_b(k^2)}
\end{equation}%%
where we have introduced the Green function
\begin{equation}
\mathcal{G}( p^2, k^2)=\frac{1}{E_p-E_k+i \epsilon}
\end{equation}
The function $A$ is the numerator of the tree-level amplitude~\eqref{Atree} with the $k$-legs  
satisfying 3-momentum (but not energy) conservation. We notice that $A(\vec{p},\vec{k}\,)$ and
$A(\vec{k},\vec{p}\,')$ can be written as
\begin{equation}
A(\vec{p},\vec{k}\,)=\tilde{A}(p^2,k^2)+B(\vec{p},\vec{k}),
\end{equation}
\begin{equation}
A(\vec{k},\vec{p}\,')=\tilde{A}(p^2,k^2)+B(\vec{p}\,',\vec{k})
\end{equation}
where $\tilde{A}$ is ${\vec q\,}$-independent and function of $|\vec{p}|=p$ and $|\vec{k}|=k$. The classical contribution
from the iterated Born amplitude is hence 
\begin{equation}
\mathcal{M}^{\rm Iterated}=-\frac{16 \pi^2 G_N^2}{E_a(p^2)E_b(p^2)} \int
\frac{d^dk}{(2\pi)^d}\frac{\mathcal{G}(p^2,k^2){Q}(p^2, k^2)}{|\vec{p}-\vec{k}\,|^2|\vec{p}\,'-\vec{k}\,|^2}
\end{equation}
where 
\begin{equation}
Q(p^2,k^2)=\frac{\tilde{A}^2(p^2,k^2)}{E_a(k^2) E_b(k^2)}
\end{equation}
We now expand $ Q$ around $p^2$,
\begin{equation}
\label{41}
 Q(p^2,k^2)= Q_{k=p}+(k^2-p^2)\partial_{k^2}Q_{k^2=p^2}+\cdots,
\end{equation}
\begin{equation}
 Q_{k^2=p^2}= \frac{\tilde{A}^2_{k^2=p^2}}{E_a(p^2)E_b(p^2)} ~=~ \frac{c_1^2}{E^2_p \xi},
\end{equation}
\begin{equation}
\partial_{{k}^2} Q_{k^2=p^2} = -\frac{1}{E^2_p\xi^2}\bigg(2c_1 p_1 \cdot p_3+\frac{c_1^2}{2E^2_p\xi}(1-2\xi) \bigg)
\end{equation}
The Green function $\mathcal{G}$ likewise admits a Laurent expansion
in $k^2$
\begin{equation}
\label{44}
\mathcal{G}(p^2, k^2)=\frac{2E_p \xi}{{p}^{\,2}-{k}^2}+\frac{3\xi-1}{2E_p \xi}+\cdots
\end{equation}
Combining terms, the Born subtraction can hence be expressed as
\begin{multline}
\mathcal{M}^{\rm Iterated}=\frac{32\pi^2G_N^2}{E_p^3 \xi}c^2_1 \int
\frac{d^dk}{(2\pi)^d}\frac{1}{|\vec{p}-\vec{k}|^2|\vec{p'}-\vec{k}|^2(k^2-p^2)}\cr
-\frac{16\pi^2G_N^2}{E_p^3\xi^2}\bigg(\frac{c^2_1(1-\xi)}{2E_p^2 \xi}+4c_1
p_1 \cdot p_3 \bigg)\int \frac{d^dk}{(2\pi)^d}\frac{1}{|\vec{p}-\vec{k}|^2|\vec{p}\,'-\vec{k}|^2}+\cdots
\end{multline}
Evaluating the remaining three-dimensional integrals, we find
\begin{equation}
\label{iter}
\mathcal{M}^{\rm Iterated} = \frac{i \pi G_N^2}{E_p^3 \xi}\frac{4c^2_1}{|\vec{p}|}\frac{(\log|\vec{q}|^2-\frac{2}{3-d})}{|\vec{q}|^2} 
 +\frac{2\pi^2G_N^2}{E_p^3\xi^2|\vec{q}|}\bigg(\frac{c^2_1(\xi-1)}{2E_p^2 \xi}-4c_1p_1 \cdot p_3 \bigg)
\end{equation}
The second-order post-Minkowskian potential in momentum space  is thus given by
\begin{equation}
V_{\rm 2PM}(p,q) = \mathcal{M}^{\rm 1-loop}+\mathcal{M}^{\rm Iterated}
\end{equation}
leading to 
\vspace{2mm}
\begin{equation}
V_{\rm 2PM}(p,q\,)= \frac{\pi^2 G_N^2}{E_p^2 \xi |\vec{q}\,|}\bigg[\frac{1}{2}\bigg(\frac{c_{\triangleright}}{m_{a}}+\frac{c_{\triangleleft}}{m_b}\bigg)  +\frac{2}{E_p\xi }\bigg(\frac{c^2_1(\xi-1)}{2E_p^2 \xi}-4c_1p_1 \cdot p_3\bigg) \bigg]
\end{equation}
\newline
or, in coordinate space,
\vspace{2mm}
\begin{equation}
V_{\rm 2PM}(p,r)=\frac{ G_N^2}{r^2}
\frac{1}{E_p^2
  \xi}\bigg[\frac{1}{4}\bigg(\frac{c_{\triangleright}}{m_{a}}+\frac{c_{\triangleleft}}{m_b}\bigg)
+\bigg(\frac{c^2_1(\xi-1)}{2E_p^3 \xi^2}-
{4 c_1 p_1 \cdot p_3\over E_p\xi} \bigg) \bigg] .
\end{equation}
This agrees with what has been previously obtained in ref.~\cite{Cheung:2018wkq}
(taking into account that $c_1$ here  is $E^2\xi$ times $c_1$ in~\cite{Cheung:2018wkq}).
As expected on physical grounds, the imaginary part which is composed of super-classical and infrared divergent pieces has cancelled, leaving a finite and well-defined post-Minkowskian potential at $d=3$. That such cancellation had to occur was expected on physical ground, since the imaginary part clearly cannot affect classical motion. Interestingly, the evaluation of the same potential in $\mathcal{N}=8$ supergravity has shown no contributions coming from triangle topologies \cite{Caron-Huot:2018ape}.

\subsection{The post-Minkowskian scattering angle}

In~\cite{Bjerrum-Bohr:2018xdl} a one-loop formula for the gravitational eikonal limit~\cite{Kabat:1992tb,Akhoury:2013yua} 
generalized to the scattering of two objects of different masses $m_a$ and $m_b$ was used to deduce the classical 
scattering angle to second post-Minkowskian order directly from the scattering amplitude. 
An alternative method based on the Hamiltonian~\cite{Damour:1988mr} has recently been revived in connection with
the third post-Minkowskian scattering amplitude calculation~\cite{Bern:2019nnu,Antonelli:2019ytb} and we here briefly summarize the method at second order in $G_N$. Since the motion lies on a plane, we can introduce the following coordinates on the phase space $(r,\phi,p_r,p_{\phi})$ so as to express the momentum in the center of mass frame as
\begin{equation}
{ p}^{\, 2}=p^2_r+\frac{L^2}{r^2} 
\end{equation}
being $L$ the conserved angular momentum of our binary system, with constant energy $E$
\begin{equation}
\sqrt{p^2+m_a^2}+\sqrt{p^2+m_b^2} +V_{\rm 1PM}(p,r)+V_{\rm 2PM}(p,r) =E
\end{equation}
This equation can be solved perturbatively in $G_N$ for
${ p}^{\, 2}={ p}^{\, 2} (E,L,r) $
\begin{equation}
 { p}^{\,2}={ p}^{\,2}_0+\frac{G_Nf_1}{r}+\frac{G^2_N f_2}{r^2}+\cdots
\end{equation}
Using $s=(p_1+p_3)^2$
\begin{equation}
\label{effe}
p^2_0=\frac{(p_1 \cdot p_3)^2-m^2_1m^2_2}{s},  \qquad
f_1=-\frac{2c_1}{\sqrt{s}}, \qquad
f_2=-\frac{1}{2\sqrt{s}}\bigg(\frac{c_{\triangleright}}{m_a}+\frac{c_{\triangleleft}}{m_b} \bigg)
\end{equation}
It is
straightforward to derive the following expression for the
change in the angular variable $\phi$ during scattering (see for
instance~\cite{Damour:1988mr,Damour:2016gwp})
\begin{equation}
\Delta \phi=\pi+ \chi(E,L),
\end{equation}
where the scattering angle is given by
\begin{equation}
\chi(E,L)=-2\int_{r_{min}}^{+\infty}dr \frac{\partial p_r}{\partial L}- \pi
\end{equation}
Here $r_{min}$ is the positive root for the condition of turning point at $p_r=0$ with
\begin{equation}
p_r=\sqrt{p^2_0-\frac{L^2}{r^2}+\frac{G_Nf_1}{r}+\frac{G^2_N f_2}{r^2}}
\end{equation}
Introducing  $r_0\equiv L/p_0$ we note that $p_r$ can be rewritten as
\begin{equation}
p_r=\frac{p_0}{r}\sqrt{r^2+r \frac{G_N f_1}{p_0^2}+\frac{G^2_Nf_2}{p^2_0}-r^2_0} 
=\frac{p_0}{r}\sqrt{r-r^{+}}\sqrt{r-r^{-}}\,,
\end{equation}
\begin{equation}
r^{\pm}=-\frac{G_Nf_1}{2p^2_0} \pm \sqrt{\frac{G^2_Nf^2_1}{4p^4_0}-\frac{G^2_N f_2}{p^2_0}+r^2_0}
\end{equation}
Since $r_{min}=r^+$, the scattering angle becomes
\begin{equation}
\chi(E,L)= 2\int_{r^{+}}^{+\infty}\frac{dr}{r}\frac{r_0}{\sqrt{(r-r^+)(r-r^{-})}} - \pi
\end{equation}
The integral so expressed can be performed analytically without the need of regularization. We get
\begin{equation}
\chi(E,L)=\frac{4 r_0}{\sqrt{-r^+ r^-}}\arccos\sqrt{\frac{r^+}{r^+-r^-}}- \pi 
\end{equation}
Taylor-expanding the scattering angle to second post-Minkowskian order we arrive at the final result
\begin{equation}
\label{sca}
\chi(E,L)=\frac{G_N f_1}{p_0 L}+\frac{G^2_Nf_2 \pi}{2L^2}+\cdots
\end{equation}
In terms of $\hat{\cal{M}}^2 \equiv s - m_a^2 - m_b^2$ and the impact parameter $b$, where $L=pb$, we have
\begin{equation}
\chi(E,b) = \frac{4G_N s}{b}\left[\frac{{\hat{{\cal{M}}}}^4 - 2m_a^2m_b^2}{{\hat{\cal{M}}}^4 - 4m_a^2m_b^2}
+ \frac{3\pi}{16}\frac{G_N(m_a+m_b)}{b}\frac{5{\hat{\cal{M}}}^4 - 4m_a^2m_b^2}{{\hat{\cal{M}}}^4 - 4m_a^2m_b^2}\right] 
 \end{equation}
which agrees with the result of~\cite{Westpfahl} at second
post-Minkowskian order. In particular, since $f_1$ and $f_2$ do not
depend on box topologies (\ref{effe}), also the scattering angle
(\ref{sca}) receives no contributions from these, a known fact from
the eikonal approach in four dimensions.
The details of the calculation based on the Hamiltonian is, on the surface, quite different from the eikonal approach. It would be
interesting to establish the precise link between the two, first identifying the precise exponentiation formula for the eikonal
limit beyond second post-Minkowskian order.

\section{Conclusion}

Using the conventional approach to determining the interaction potential in perturbative gravity we have demonstrated that
it can be extended to the relativistic setting by means of a one-particle Hamiltonian and
associated Salpeter equation. We have used the Lippmann-Schwinger equation to derive straightforwardly the needed Born
subtractions at arbitrary loop order. The resulting Fourier-transformed 
post-Minkowskian Hamiltonian
\begin{equation}
\mathcal{H}_{\rm 2PM}(p,r)=\sqrt{p^2+m_a^2}+\sqrt{p^2+m^2_b}+V_{\rm 1PM}(p,r)+V_{\rm 2PM}(p,r)\,,
\end{equation}
agrees with the one derived in ref.~\cite{Cheung:2018wkq} based on an effective field theory expansion in operators that
can contribute to the given order, supplemented with the matching condition that the scattering amplitude as computed in
the effective theory agrees with the one computed from the full one-loop expression of the Einstein-Hilbert action (plus
scalars). 

The resulting post-Minkowskian Salpeter equation is not an effective low-energy theory (momentum is not limited), 
but rather a small $|\vec{q}\,|/m$ approximation where small momentum is exchanged and only particle states are summed over. 
It is encouraging that preliminary
results indicate that the corresponding two-loop Hamiltonian~\cite{Bern:2019nnu} may improve the computation
of two-body dynamics as compared to the conventional post-Newtonian expansion for bound states~\cite{Antonelli:2019ytb}.
The post-Minkowskian Hamiltonian also appears to provide a short-cut towards computing the scattering angle without
first demonstrating exponentiation (and potential correction terms) as in the eikonal approach. It would be interesting
to demonstrate the equivalence between those two scattering angle computations in all generality.

%%%%%%%%%%%%%%%%%%%%%
\begin{acknowledgements}
  This work has been based partly on funding from the European Union's
  Horizon 2020 research and innovation programme under the Marie
  Sk\l{}odowska-Curie grant agreement No. 764850 (``SAGEX''). The work of
  N.E.J.B.-B. and P.H.D. was supported in part by the Danish
  National Research Foundation (DNRF91).  N.E.J.B.-B. in addition
  acknowledges partial support from the Carlsberg Foundation. The
  research of P.V. has received funding the ANR grant ``Amplitudes'' ANR-17-
  CE31-0001-01, and is partially supported by Laboratory of Mirror
  Symmetry NRU HSE, RF Government grant, ag. N${}^\circ$
  14.641.31.0001.  P.V. also thanks the Galileo Galilei Institute for
  Theoretical Physics and INFN for hospitality and partial support
  during the workshop "String Theory from a worldsheet perspective"
  where part of this work has been done.
\end{acknowledgements}

\end{document}